\documentstyle[twocolumn,aps,floats,amssymb,epsfig]{revtex}

\begin{document}

\def\d{{\rm d}}
\def\e{{\rm e}}
\def\i{{\rm i}}
\def\O{{\rm O}}
\def\half{\mbox{$\frac12$}}
\def\eref#1{(\protect\ref{#1})}
\def\etal{{\it{}et~al.}}
\def\Li{\mathop{\rm Li}}
\def\av#1{\left\langle#1\right\rangle}
\def\set#1{\left\lbrace#1\right\rbrace}
\def\stirling#1#2{\Bigl\lbrace{#1\atop#2}\Bigr\rbrace}

\draft
\tolerance = 10000

\renewcommand{\topfraction}{0.9}
\renewcommand{\textfraction}{0.1}
\renewcommand{\floatpagefraction}{0.9}
\setlength{\tabcolsep}{4pt}

%Fixing abstract in twocolumn mode
\twocolumn[\hsize\textwidth\columnwidth\hsize\csname @twocolumnfalse\endcsname

\title{Network robustness and fragility:
Percolation on random graphs}
\author{Duncan S. Callaway$^{1}$, M. E. J. Newman$^{2,3}$, Steven H.
  Strogatz$^{1,2}$, and Duncan J. Watts$^{4}$}
\address{$^1$Department of Theoretical and Applied Mechanics,
Cornell University, Ithaca, NY 14853--1503}
\address{$^2$Center for Applied Mathematics, Cornell University,
Ithaca, NY 14853--3801}
\address{$^3$Santa Fe Institute, 1399 Hyde Park Road, Santa Fe, NM 87501}
\address{$^4$Department of Sociology, Columbia University,
1180 Amsterdam Avenue, New York, NY 10027}
\maketitle

\begin{abstract}
  Recent work on the internet, social networks, and the power grid has
  addressed the resilience of these networks to either random or targeted
  deletion of network nodes.  Such deletions include, for example, the
  failure of internet routers or power transmission lines.  Percolation
  models on random graphs provide a simple representation of this process,
  but have typically been limited to graphs with Poisson degree
  distribution at their vertices.  Such graphs are quite unlike real world
  networks, which often possess power-law or other highly skewed degree
  distributions.  In this paper we study percolation on graphs with
  completely general degree distribution, giving exact solutions for a
  variety of cases, including site percolation, bond percolation, and
  models in which occupation probabilities depend on vertex degree.  We
  discuss the application of our theory to the understanding of network
  resilience.
\end{abstract}

%\pacs{64.60.Ak, 87.23.Ge, 05.70.Jk}

%Fixing abstract in twocolumn mode
]

\newpage

The internet, airline routes, and electric power grids are all examples of
networks whose function relies crucially on the pattern of interconnection
between the components of the system.  An important property of such
connection patterns is their robustness---or lack thereof---to removal of
network nodes~\cite{AJB00}, which can be modeled as a percolation process
on a graph representing the network~\cite{CEBH00}.  Vertices on the graph
are considered occupied or not, depending on whether the network nodes they
represent (routers, airports, power stations) are functioning normally.
Occupation probabilities for different vertices may be uniform, or may
depend on, for example, the number of connections they have to other
vertices, also called the vertex degree.  Then we observe the properties of
percolation clusters on the graph, particularly their connectivity, as the
function determining occupation probability is varied.  Previous results on models of this
type~\cite{AJB00,CEBH00,Broder00} suggest that, if the connection patterns
are chosen appropriately, the network can be made highly resilient to
random deletion of nodes, although it may be susceptible to an ``attack''
which specifically targets nodes of high degree.  We can also consider bond
percolation on graphs as a model of robustness of networks to failure of
the links between nodes (e.g.,~fiber optic lines, power transmission
cables, and so forth), or combined site and bond percolation as a model of
robustness against failure of either nodes or links.

Percolation models built on networks have also been used to model the
spread of disease through communities~\cite{BMST97,NW99}.  In such models a
node in the network represents a potential host for the disease, and is
occupied if that host is susceptible to the disease.  Links between nodes
represent contacts capable of transmitting the disease between individuals
and may be occupied with some prescribed probability to represent the
fraction of such contacts which actually result in transmission.  A
percolation transition in such a model represents the onset of an epidemic.
Similar models can be used to represent the propagation of computer
viruses~\cite{KW91}.

The simplest and most widely studied model of undirected networks is the
random graph~\cite{Bollobas85}, which has been investigated in depth for
several decades now.  However, random graphs suffer (at least) one serious
shortcoming.  As pointed out by a number of
authors~\cite{Broder00,FFF99,HA99,AJB99,ASBS00}, vertex degrees have a
Poisson distribution in a random graph, but real-life degree distributions
are strongly non-Poisson, often taking power-law, truncated power-law, or
exponential forms.  This has prompted researchers to study the properties
of generalized random graphs which have non-Poisson degree
distributions~\cite{MR9598,ACL00,NSW00}.

In this paper we employ the generating function formalism of
Newman~\etal~\cite{NSW00} to find exact analytic solutions for site
percolation on random graphs with any probability distribution of vertex
degree, where occupation probability is an arbitrary function of vertex
degree.  For the special case of constant occupation probability, we also
give solutions for bond and joint site/bond percolation.  Our results
indicate how robust networks should be to random deletion of vertices or
edges, or to the preferential deletion of vertices with particular degree.

We start by examining site percolation for the general case in which
occupation probability is an arbitrary function of vertex degree.  Let
$p_k$ be the probability that a randomly chosen vertex has degree~$k$, and
$q_k$ be the probability that a vertex is occupied given that it has
degree~$k$.  Then $p_k q_k$ is the probability of having degree $k$ and
being occupied, and
\begin{equation}
\label{G0_eq}
F_0(x)=\sum_{k=0}^{\infty} p_k q_k x^k
\end{equation}
is the probability generating function for this distribution~\cite{Wilf94}.
(Generating functions of this form were previously used by Watts~\cite{Watts00}
to study cascading failures in networks.)  Note that $F_0(1)=q$, where $q$
is the overall fraction of occupied sites.  If we wish to study the special
case of uniform occupation probability---ordinary site percolation---we
simply set $q_k=q$ for all~$k$.

If we follow a randomly chosen edge, the vertex we reach has degree
distribution proportional to $k p_k$ rather than just $p_k$ because a
randomly chosen edge is more likely to lead to a vertex of higher degree.
Hence the equivalent of~\eref{G0_eq} for such a vertex is~\cite{NSW00}
\begin{equation}
F_1(x)=\frac{\sum_k kp_kq_kx^{k-1}}{\sum_k kp_k} = \frac{F_0'(x)}{z},
\end{equation}
where $z$ is the average vertex degree.

Now let $H_1(x)$ be the generating function for the probability that one
end of a randomly chosen {\em edge\/} on the graph leads to a percolation
cluster of a given number of occupied vertices.  The cluster may contain
zero vertices if the vertex at the end of the edge in question is
unoccupied, which happens with probability~$1-F_1(1)$, or the edge may lead
to an occupied vertex with a number $k$ of other edges leading out of it,
distributed according to $F_1(x)$.  This means that $H_1(x)$ satisfies a
self-consistency condition of the form~\cite{NSW00,MN00,note3}

\begin{equation} 
\label{H1_eq} 
H_1(x) = 1 - F_1(1) + x F_1(H_1(x)).
\end{equation} 
The probability distribution for the size of the cluster to which a
randomly chosen {\em vertex\/} belongs is similarly generated by $H_0(x)$,
where
\begin{equation} 
\label{H0_eq} 
H_0(x) = 1 - F_0(1) + x F_0(H_1(x)).
\end{equation} 
Together, Eqs.~(\ref{G0_eq}--\ref{H0_eq}) determine the cluster size
distribution for site percolation on a graph of arbitrary degree
distribution.  From these equations we can determine several quantities of
interest such as mean cluster size, position of the percolation threshold,
and giant component size, as demonstrated below.

For the special case of uniform (degree-independent) site occupation
probability, $q_k=q$ for all~$k$, Eqs.~\eref{H1_eq} and~\eref{H0_eq}
simplify to
\begin{eqnarray}
\label{sitep_eq}
H_1(x) &=& 1 - q + q x G_1(H_1(x)),\\
\label{sitep2_eq}
H_0(x) &=& 1 - q + q x G_0(H_1(x)),
\end{eqnarray}
where $G_0(x)=\sum_k{p_kx^k}$ and $G_1(x)=G_0'(x)/z$ are the generating
functions for vertex degree alone introduced in Ref.~\onlinecite{NSW00}.
For bond percolation with uniform occupation probability, we find that
\begin{equation}
\label{bondp_eq}
H_0(x) = x G_0(H_1(x)),
\end{equation}
with $H_1(x)$ given by Eq.~\eref{sitep_eq} again, and for joint site/bond
percolation with uniform site and bond occupation probabilities $q_s$
and~$q_b$, we have
\begin{eqnarray}
H_1(x) &=& 1 - q_s q_b + q_s q_b x G_1(H_1(x)),\\
H_0(x) &=& 1 - q_s + q_s x G_0(H_1(x)), 
\end{eqnarray} 
and indeed Eqs.~(\ref{sitep_eq}--\ref{bondp_eq}) may be considered special
cases of these last two equations when either $q_s$ or $q_b$ is~1.

We now apply these results to the study of network robustness in a variety
of cases.  First, we consider the case of uniform site occupation
probability embodied in Eqs.~\eref{sitep_eq} and~\eref{sitep2_eq}, which
corresponds to random removal of nodes from a network, for example through
failure of routers in a data network, or through random vaccination of a
population against a disease.

Typically, no closed-form solution exists for Eq.~\eref{sitep_eq}, but it
is possible to determine the terms of $H_1(x)$ to any finite order $n$ by
iterating Eq.~\eref{sitep_eq} $n+1$ times starting from an initial value of
$H_1=1$.  The probability distribution of cluster sizes can then be
calculated exactly by substituting into Eq.~\eref{H0_eq} and expanding
about~$x=0$.  To test this method, we have performed
simulations~\cite{note1} of site percolation on random graphs with vertex
degrees distributed according to the truncated power law
\begin{equation}
\label{pld_eq}
p_k = \left\lbrace \begin{array}{l@{\quad}l}
0                        & \mbox{for $k=0$} \\
Ck^{-\tau}\e^{-k/\kappa} & \mbox{for $k\ge1$.}
                   \end{array} \right.
\end{equation}
Our reasons for choosing this distribution are two-fold.  First, it is seen
in a number of real-world social networks including collaboration networks
of movie actors~\cite{ASBS00} and scientists~\cite{Newman00}.  The pure
power-law distributions seen in internet data~\cite{FFF99,HA99,AJB99} are
also included in~\eref{pld_eq} as a special case~$\kappa\to\infty$.
Second, the distribution has technical advantages over a pure power-law
form
because the exponential cutoff regularizes the calculations, so that the
generating functions and their derivatives are finite.  For pure power-law
forms on the other hand, the calculations diverge, indicating that
real-world networks cannot take a pure power-law form and must have some
cutoff (presumably dependent on the system size).

\begin{figure}
\begin{center}
\psfig{figure=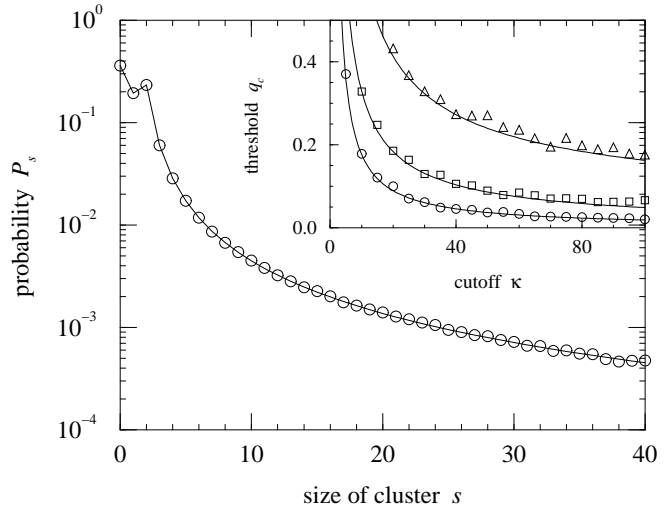,width=\columnwidth}
\end{center}
\caption{Probability $P_s$ that a randomly chosen vertex belongs to a
cluster of $s$ sites for $\kappa=10$, $\tau=2.5$, and $p=0.65$ from
numerical simulation on systems of $10^7$ sites (circles) and our exact
solution (solid line).  Inset: the percolation threshold $q_c$ from
Eq.~\eref{pc_eq} (solid lines), versus computer simulations with
$\tau=1.5$ (circles), $2.0$ (squares), and $2.5$ (triangles).}
\label{csd_fig}
\end{figure}

Figure~\ref{csd_fig} shows the cluster size distribution from our
simulations along with the exact solution for the same values from the
generating function formalism.  The agreement between the two is good.

The sizes of the clusters correspond, for instance, to the sizes of
outbreaks of a disease among groups of susceptible individuals.  The
parameter values used in Fig.~\ref{csd_fig} are below the percolation
threshold for this particular degree distribution, and hence all outbreaks
are small and there is no epidemic behavior.  The mean cluster size is
\begin{equation}
\av{s} = H_0'(1) = q + q G_0'(1) H_1'(1)
       = q\left[1+\frac{qG_0'(1)}{1-qG_1'(1)}\right],
\end{equation}
which diverges when $1-qG_1'(1)=0$.  This point marks the percolation
threshold of the system, the point at which a giant component of connected
vertices first forms.  Thus the critical occupation probability is
\begin{equation}
\label{pc_eq}
q_c=\frac{1}{G_1'(1)}.
\end{equation}
A result equivalent to this one has been derived previously by
Cohen~\etal~\cite{CEBH00} by different means.

In the language of disease propagation $q_c$ is the point at which an
epidemic of the disease first occurs.  In the language of network
robustness, it is the point at which the network achieves large scale
connectivity, and can therefore function as an effective distribution
network.  Conversely, if we are approaching the transition from values of
$q$ above $q_c$ it is the point at which a sufficient number of individuals
are immune to a disease to prevent it from spreading, or the point at which
a large enough number of nodes have been deleted from a distribution
network to prevent distribution on large-scales.

The inset of Fig.~\ref{csd_fig} shows the behavior of the percolation
threshold with the cutoff parameter $\kappa$ for a variety of values
of~$\tau$.  Note that as the values of $\kappa$ become large, the
percolation threshold becomes small, indicating a high degree of robustness
of the network to random deletion of nodes.  For $\tau=2.5$ (roughly the
exponent for the internet data \cite{FFF99}) and $\kappa=100$, the
percolation threshold is $q_c=0.17$, indicating that one can remove more
than 80\% of the nodes in the network without destroying the giant
component---the network will still possess large-scale connectivity.  This
result agrees with recent studies of the internet~\cite{AJB00,CEBH00} which
indicate that network connectivity should be highly robust against the
random removal of nodes.

\begin{figure}
\begin{center}
\psfig{figure=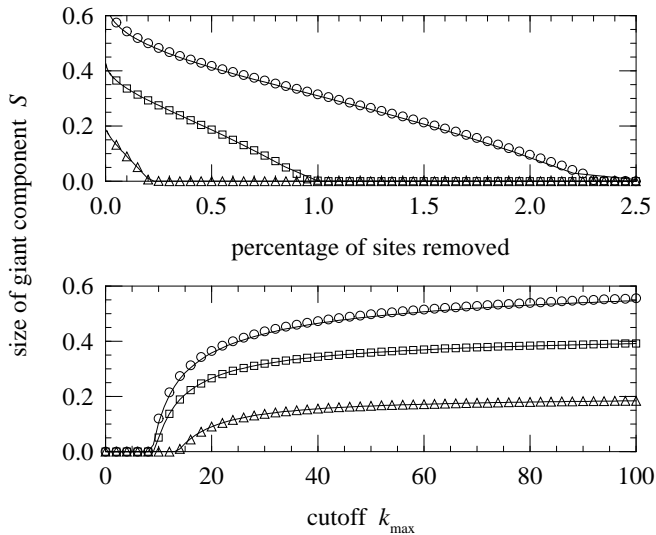,width=\columnwidth}
\end{center}
\caption{Size of the giant component $S$ in graphs with power-law degree
distribution and all vertices with degree greater than $k_{\rm max}$
unoccupied, for $\tau=2.4$ (circles), $2.7$ (squares), and $3.0$
(triangles).  Points are simulation results for systems with $10^7$
vertices, solid lines are the exact solution.  Upper frame: as a function
of fraction of vertices unoccupied.  Lower frame: as a function of the
cutoff parameter $k_{\rm max}$.}
\label{gcsize_fig}
\end{figure}

Another issue that has attracted considerable recent attention is the
question of robustness of a network to non-random deletion targeted
specifically at nodes with high degree.  Albert~\etal~\cite{AJB00} and
Broder~\etal~\cite{Broder00} both looked at the connectivity of a network
with power-law distributed vertex degrees as the vertices with highest
degree were progressively removed.  In the language of our percolation
models, this is equivalent to setting
\begin{equation}
q_k = \theta(k_{\rm max}-k),
\end{equation}
where $\theta$ is the Heaviside step-function~\cite{note2}.  This removes
(unoccupies) all vertices with degree greater than $k_{\rm max}$.  To
investigate the effect of this removal, we calculate the size of the giant
component in the network, if there is one.  Above the percolation
transition the generating function $H_0(x)$ gives the distribution of the
sizes of clusters of vertices which are {\em not\/} in the giant
component~\cite{MN00}, which means that $H_0(1)$ is equal to the fraction
of the graph which is not occupied by the giant component.  The fraction
$S$ which {\em is\/} occupied by the giant component is therefore given by
\begin{equation}
S = 1 - H_0(1) = F_0(1) - F_0(u),
\label{S_eq}
\end{equation}
where $u$ is a solution of the self-consistency condition
\begin{equation}
u = 1 - F_1(1) + F_1(u).
\end{equation}
In cases where this last equation is not exactly solvable we can evaluate
$u$ by numerical iteration starting from a suitable initial value.  In
Fig.~\ref{gcsize_fig} we show the results for $S$ from this calculation for
graphs with pure power-law degree distributions as a function of $k_{\rm
max}$ for a variety of values of $\tau$.  (The removal of vertices with
high degree regularizes the calculation in a similar way to the inclusion
of the cutoff $\kappa$ in our earlier calculation, so no other cutoff is
needed in this case.)  On the same plot we also show simulation results for
this problem, and once more agreement of theory and simulation is good.

Opinions appear to differ over whether networks such as this are robust or
fragile to this selective removal of vertices.  Albert~\etal~\cite{AJB00}
point out that only a small fraction of the highest-degree vertices need be
removed to destroy the giant component in the network and hence remove all
long-range connectivity.  Conversely, Broder~\etal~\cite{Broder00} point
out that one can remove all vertices with degree greater than $k_{\rm max}$
and still have a giant component even for surprisingly small values
of~$k_{\rm max}$.  As we show in Fig.~\ref{gcsize_fig}, both viewpoints are
correct: they are merely different representations of the same data.  In
the upper frame of the figure, we plot giant component size as a function
of the fraction of vertices removed from the network, and it is clear that
the giant component disappears when only a small percentage are
removed---just 1\% for the case $\tau=2.7$---so that the network appears
fragile.  In the lower frame we show the same data as a function of $k_{\rm
max}$, the highest remaining degree vertex, and we see that when viewed in
this way the network is, in a sense, robust, since $k_{\rm max}$ must be
very small to destroy the giant component completely---just 10 in the case
of $\tau=2.7$.

To conclude, we have used generating function methods to solve exactly for
the behavior of a variety of percolation models on random graphs with any
distribution of vertex degrees, including uniform site, bond and site/bond
percolation, and percolation in which occupation probability is a function
of vertex degree.  Percolation systems on graphs such as these have been
suggested as models for the robustness of communication or distribution
networks to breakdown or sabotage, and for the spread of disease through
communities possessing some resistance to infection.  Our exact solutions
allow us to make predictions about the behavior of such model systems under
quite general types of breakdown or interference.  Among other results, we
find that a distribution network such as the internet, which has an
approximately power-law vertex degree distribution, should be highly robust
against random removal of nodes (for example, random failure of routers),
but is relatively fragile, at least in terms of fraction of nodes removed,
to the specific removal of the most highly connected nodes.

The authors would like to thank Jon Kleinberg for illuminating
conversations.  This work was funded in part by the National Science
Foundation, the Electric Power Research Institute, and the Army Research
Office.


\begin{references}
\bibitem{note3} The derivation of this equation assumes that all clusters
other than the giant component contain no loops.  It is straightforward to
demonstrate that this assumption is exact in the limit of large system size
for degree distributions with finite mean.  See
Refs.~\onlinecite{Bollobas85}, \onlinecite{MR9598}, and~\onlinecite{ACL00}
for discussions of this point.
\bibitem{note2} In fact, the approaches of Albert~\etal~\cite{AJB00} and
Broder~\etal~\cite{Broder00} differ slightly.  Broder~\etal\ simply removed
the highest degree vertices from the network, whereas Albert~\etal\
recalculated vertex degrees after the removal of each vertex and its
associated edges, and then removed the next highest degree vertex.  Our
calculations are equivalent to the method of Broder~\etal, although in
practice there appears to be little difference in qualitative behavior
between the two.
\end{references}
\end{document}